\documentclass[lettersize,journal]{IEEEtran}
\usepackage{amsmath,amsfonts}
\usepackage{algorithmic}
\usepackage{algorithm}
\usepackage{array}
\usepackage[caption=false,font=normalsize,labelfont=sf,textfont=sf]{subfig}
\usepackage{textcomp}
\usepackage{stfloats}
\usepackage{url}
\usepackage{verbatim}
\usepackage{graphicx}
\usepackage{cite}
\usepackage{booktabs}

\usepackage{booktabs} 

\usepackage{tikz}
\usepackage{yquant,tikz}
\usepackage{xcolor}
\usepackage{amsfonts}  
\usepackage{float} 
\usetikzlibrary{quantikz}
\usetikzlibrary{shapes.geometric, arrows, positioning}
\tikzset{
    block/.style = {rectangle, draw, text width=6em, text centered, rounded corners, minimum height=4em},
    line/.style = {draw, -latex'}
}
\usetikzlibrary{positioning}
\makeatletter
\DeclareRobustCommand\rvdots{%
\vbox{%
\baselineskip4\p@\lineskiplimit\z@%
\kern-\p@%
\hbox{.}\hbox{.}\hbox{.}%
}%
}

\begin{document}

\title{Quantum Neural Networks for Solving Power System Transient Simulation Problem}

\author{Mohammadreza Soltaninia, Junpeng Zhan,~\IEEEmembership{Member,~IEEE}
\thanks{The authors are with the Department of Electrical Engineering, Alfred University, Alfred, NY, USA. This work is funded by NSF under the award \# 2138702.}
}

\markboth{}
{Shell \MakeLowercase{\textit{et al.}}: A Sample Article Using IEEEtran.cls for IEEE Journals}


\maketitle

\begin{abstract}
Quantum computing, leveraging principles of quantum mechanics, represents a transformative approach in computational methodologies, offering significant enhancements over traditional classical systems. This study tackles the complex and computationally demanding task of simulating power system transients through solving differential algebraic equations (DAEs). We introduce two novel Quantum Neural Networks (QNNs): the Sinusoidal-Friendly QNN and the Polynomial-Friendly QNN, proposing them as effective alternatives to conventional simulation techniques. Our application of these QNNs successfully simulates two small power systems, demonstrating their potential to achieve good accuracy. We further explore various configurations, including time intervals, training points, and the selection of classical optimizers, to optimize the solving of DAEs using QNNs. This research not only marks a pioneering effort in applying quantum computing to power system simulations but also expands the potential of quantum technologies in addressing intricate engineering challenges.
\end{abstract}

\begin{IEEEkeywords}
Differential-algebraic equations, power system transient simulation, quantum neural network.
\end{IEEEkeywords}

\section{Introduction}
\IEEEPARstart{Q}{uantum} computing represents a paradigm shift in computational capabilities, leveraging principles of quantum mechanics to process information in fundamentally different ways from classical computing. Traditional computers use binary bits as the basic unit of data, which can either be a 0 or a 1. In contrast, quantum computers use quantum bits, or qubits, which can exist in multiple states simultaneously, thanks to the phenomena of superposition and entanglement \cite{Nielsen2011}. This allows quantum computers to process a vast number of possibilities concurrently, offering potential exponential speed-ups for certain computational tasks, for example, in factoring large integer numbers \cite{Shor1994}, in searching unstructured database \cite{Grover1996, Grover1997}.

Recent advancements in quantum hardware \cite{Madsen2022, Hyyppa2022, Arrazola2021} and the development of sophisticated algorithms \cite{Lloyd2020NDE, JordanZoo, Bharti2021, Zhan2022VQS, Biamonte2017} have made quantum computing a highly promising area of research. These advancements make quantum computing a promising field for a variety of applications, particularly in areas that require handling complex, high-dimensional datasets or simulations where classical computers struggle.

One of the pivotal developments in this field is the Quantum Neural Network (QNN) \cite{Liao2022, Skolik2021}. QNNs integrate the principles of quantum computing with the architectural concepts of classical neural networks. This hybrid approach has been explored for various applications, including drug discovery, financial modeling, and, notably, complex system simulations such as those needed in power systems.

Quantum computing has been explored for various power system applications, including power flow analysis \cite{Feng2021PF}, unit commitment \cite{Feng2023UC}, system reliability \cite{Nikmehr2023Reliability}, and stability assessment \cite{Yifan2023Noise}. There is a burgeoning body of literature, as indicated by several review papers \cite{Gao2023Review, Golestan2023Review, Zhou2022Review}, that discusses the integration of quantum computing into power system operations, underscoring the participation of entities such as the Department of Energy (DOE) and various utility companies in these research efforts.

Simulation of power systems is of paramount importance for ensuring their stability, efficiency, and reliability \cite{Chow2020PowerSystem, Barret2024, Sun2023book, FanMiao2024}, particularly as the integration of renewable energy sources like wind and solar continues to expand. These renewables are incorporated into power systems through power electronic devices, such as inverters, which necessitate the simultaneous handling of both small and large time steps in simulations. This is due to the rapid responses of the power electronic devices contrasted with the slower-moving dynamics of synchronous generators. Accurate modeling of these diverse dynamics is essential for effective grid management and planning, highlighting the critical need for robust simulation tools capable of managing complex interactions across varying time scales.

The simulation of power systems typically involves solving Differential-Algebraic Equations (DAEs), which describe both the dynamic behavior and the algebraic constraints within the electrical network. Traditional approaches for these simulations include numerical methods such as Euler’s and Runge-Kuutta methods \cite{DE2}, which discretize the equations to approximate solutions, and semi-analytical methods \cite{Sun2023book,Liu2020}, which combine analytical and numerical techniques for improved accuracy. More recently, Physics-Informed Neural Networks (PINNs) \cite{pinn} have been applied to solve differential equations (DEs) by training neural networks to adhere to the underlying physics of the systems, thus providing a data-driven approach to simulation. 

However, each of these methods has its limitations. Numerical and semi-analytical methods can become computationally intensive and may not scale efficiently with the increasing complexity of modern power systems. PINNs, while innovative, often require extensive data and computational resources, as they rely on classical neural networks that utilize a large number of parameters to capture complex functions.

In this context, QNNs present a promising alternative. QNNs leverage the principles of quantum mechanics to perform computations, utilizing a smaller number of parameters than classical neural networks to represent non-linear functions \cite{Liao2022}. This reduction in parameters can potentially lead to more efficient simulations, as QNNs could process complex computations faster and with higher precision due to quantum superposition and entanglement.

Quantum computing has been applied to solve DEs and Partial Differential Equations (PDEs) \cite{Oz2023}. However, its potential to solve DAEs remains largely unexplored. Therefore, this paper takes a pioneering step by investigating the use of QNNs to address the challenges in power system simulation. This exploration is the first of its kind and aims to demonstrate how quantum-enhanced computational models can solve the simulation of complex power systems. 

The rest of the paper is organized as follows. Section \ref{sec:problem_description} provides a description of the DAEs for the two small power systems simulated. Section \ref{sec:QNN_for_DAE} offers a general overview of the QNN. Section \ref{sec:QNN_for_simulation} details the two types of QNNs used to solve the power system simulation problems. Section \ref{sec:result} presents the simulation results. Finally, conclusions are given in Section \ref{sec:conclusion}.

\section{Problem Description}
\label{sec:problem_description}
In this section, we describe the Ordinary Differential Equations (ODEs) and DAEs of two power systems, which are utilized for the simulation studies presented in Section \ref{sec:result}.

\subsection{Single Machine Infinite Bus (SMIB) System}
This subsection details the ODE for the SMIB system, as described below \cite{Sauer2017power}:

\begin{equation}
    \frac{d\delta}{dt} = \omega - \omega_s \label{eq:ode1} 
\end{equation}
\begin{equation}
    \frac{d(\omega - \omega_s)}{dt} = K_1 - K_2 \sin(\delta) - K_3 (\omega - \omega_s) \label{eq:ode2}
\end{equation}
where $K_1 = \frac{\omega_s}{2H}T_m^0$, $K_2 = \frac{\omega_s}{2H} \left(\frac{E_cV}{X}\right)$, $K_3 = \frac{\omega_s}{2H}D$, $E_c$ is equal to the magnitude of the internal voltage of the machine, $X$ is the sum of reactance, and $T_m^0$ is the constant mechanical torque. 
Initial $\delta_0$ is $-1$, and the initial speed difference $(\omega_0 - \omega_s)$ is $7$. $K_1 = 5$, $K_2 = 10$, $K_3 = 1.7$. These data are taken from \cite{Sauer2017power}. 

We combine Eqs.~\ref{eq:ode1} and \ref{eq:ode2} into a single DE by eliminating $\omega$:
\begin{equation}
    \frac{d^2\delta}{dt^2} = K_1 - K_2 \sin(\delta) - K_3 \frac{d\delta}{dt} \label{eq:combined}
\end{equation}

\subsection{WSCC 3-machine System}


Here we provide the DAEs of the WSCC 3-machine 9-bus power system.
The DAE of the $i^{\text{th}}$ generator of the power system is given below \cite{Wang2021DAE}:
\begin{equation}
    \frac{d\delta_i}{dt} = \omega_s \Delta\omega_i \label{eq:dae1}
\end{equation}
\begin{equation}
    \frac{d\Delta\omega_i}{dt} = \frac{1}{2H_i} (P_{mi} - P_{ei} - D_i \Delta\omega_i) \label{eq:dae2}
\end{equation}
\begin{equation}
    P_{ei} = e_{xi}i_{xi} + e_{yi}i_{yi} \label{eq:dae3}
\end{equation}


\begin{equation}
    \begin{bmatrix}
        e_{xi}^{\prime} \\
        e_{yi}^{\prime}
    \end{bmatrix}
    =
    \begin{bmatrix}
        \sin\delta_i & \cos\delta_i \\
        -\cos\delta_i & \sin\delta_i
    \end{bmatrix}
    \begin{bmatrix}
        0 \\
        e_{qi}^{\prime}
    \end{bmatrix}
\end{equation}
\begin{equation}
    I_t = \begin{bmatrix}
        i_{x1} \\
        i_{y1} \\
        \vdots \\
        i_{xn} \\
        i_{yn}
    \end{bmatrix}
    =
    Y
    \begin{bmatrix}
        e_{x1}^{\prime} \\
        e_{y1}^{\prime} \\
        \vdots \\
        e_{xn}^{\prime} \\
        e_{yn}^{\prime}
    \end{bmatrix}
\end{equation}
\begin{equation}
    \begin{bmatrix}
        e_{xi} \\
        e_{yi}
    \end{bmatrix}
    =
    \begin{bmatrix}
        e_{qi}^{\prime}\cos\delta_i \\
        e_{qi}^{\prime}\sin\delta_i
    \end{bmatrix}
    -
    \begin{bmatrix}
        R_{ai} & -X_{di}^{\prime} \\
        X_{di}^{\prime} & R_{ai}
    \end{bmatrix}
    \begin{bmatrix}
        i_{xi} \\
        i_{yi}
    \end{bmatrix}
\end{equation}
where 
$n$ is the total number of generators, 
\(\delta_i\) and \(\Delta\omega_i\) are the rotor angle and rotor speed deviation from the nominal value of generator \(i\), respectively, 
\(H_i\) and \(D_i\) are the inertia and damping constants of generator \(i\), respectively, 
\(P_{mi}\) is the mechanical power of generator \(i\), 
\(P_{ei}\) is the electric power of generator \(i\) 
\(e_{xi}^{\prime}\) and \(e_{yi}^{\prime}\) are the internal bus voltages of generator \(i\) in the non-rotating coordinate, respectively,
\(e_{qi}^{\prime}\) is the field voltage of generator \(i\), 
\(i_{xi}\) and \(i_{yi}\) are the terminal currents of generator \(i\), 
\(Y\) is the admittance matrix, 
\(R_{ai}\) and \(X_{di}^{\prime}\) are the source impedance of generator \(i\), respectively,
\(e_{xi}\) and \(e_{yi}\) are the terminal voltages for the \(x\) and \(y\) axes of generator \(i\), respectively.

Initial values for \(\delta_i\), \(\omega_i\), \(\omega_s\), \(H_i\), \(P_{mi}\), \(D_i\), \(e_{qi}^{\prime}\), \(\delta_i\), \(R_{ai}\), \(X_{di}^{\prime}\) are obtained from the machine dataset given in \cite{Wang2021DAE}. Utilizing this data, the admittance matrix \(Y\) is calculated. Additionally, \(P_{ei}\) is derived from the network equations based on these initial values. 


\begin{table}[ht]
\centering
\caption{Parameter data of the three machines in the WECC system}
\label{tab:wecc_params}
\label{table:3machine_data}
\begin{tabular}{@{}lccc@{}}
\toprule
\textbf{Parameter} & \textbf{Generator 1} & \textbf{Generator 2} & \textbf{Generator 3} \\
\midrule
H & 23.64 & 6.40 & 3.01 \\
D & 23.64 & 6.40 & 3.01 \\

$R_a$ & 0 & 0 & 0 \\
$X'_d$ & 0.0608 & 0.1198 & 0.1813 \\
\midrule  
$P_m$ & 0.7164 & 1.6300 & 0.8500 \\
$e'_q$ & 1.0566 & 1.0502 & 1.0170 \\
$\delta(0)$ & 0.0626 & 1.0567 & 0.9449 \\
$\Delta e(0)$ & 0 & 0 & 0 \\
Real$(I_t)$ & 0.6889 & 1.5799 & 0.8179 \\
Imag$(I_t)$ & -0.2601 & 0.1924 & 0.1730 \\
$I_d$ & 0.2872 & 0.3523 & 0.0178 \\
$I_q$ & 0.6780 & 1.5521 & 0.8358 \\
\bottomrule
\end{tabular}
\end{table}

For the disturbance simulation, we simulate a three-phase fault near bus 7 at the end of line 5-7, which is cleared within five cycles (0.083s) by opening line 5-7. The simulation includes three distinct operational phases: pre-fault, fault-on, and post-fault conditions. The pre-fault phase spans from 0 to 10 seconds, the fault-on phase from 10.0 to 10.083 seconds, and the post-fault phase from 10.083 to 20 seconds. Each phase initializes with different starting points: pre-fault initial values are sourced from machine data to calculate \(P_{ei}\); fault-on initial values are derived from the terminal values of the pre-fault simulation at 10 seconds, with adjustments made to accommodate the new fault-on \(Y\) matrix; and post-fault initial values are taken from the end of the fault-on phase at 10.083 seconds, incorporating modifications for a subsequent \(Y\) matrix change. The respective \(Y\) matrices for each condition are documented in \cite{Wang2021DAE}.

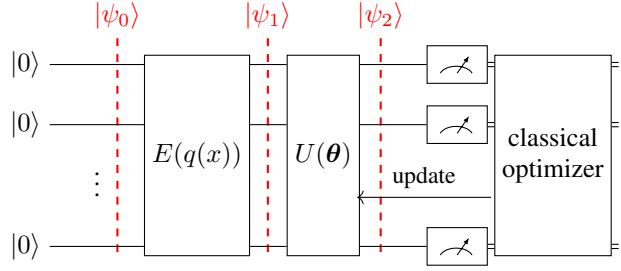
\begin{figure*}[!h]
	\centering
	\begin{tikzpicture}[label distance=4mm]
		\begin{yquant}[register/separation=3mm]
			qubit {$\ket0$} a;
			qubit {$\ket0$} b;
			nobit c;
			qubit {$\ket0$} d;
			hspace {2.8mm} a;
			hspace {2.8mm} b;
			hspace {2.8mm} c;
			hspace {2.8mm} d;
			text {$\rvdots$} c;

			\draw [red, thick, dashed] (0.9, 0.1) -- (0.9, -2.8);
			\node [red] at (0.9, 0.4) {$|\psi_0\rangle$};
			
			hspace {2.8mm} a;
			hspace {2.8mm} b;
			hspace {2.8mm} c;
			hspace {2.8mm} d;
			box {$E(q(x))$} (a,b,c,d);

			\draw [red, thick, dashed] (2.9, 0.1) -- (2.9, -2.8);
			\node [red] at (2.9, 0.4) {$|\psi_1\rangle$};
			
			hspace {4mm} a;
			hspace {4mm} b;
			hspace {4mm} c;
			hspace {4mm} d;
			box {$U(\boldsymbol{\theta})$} (a,b,c,d);
			
			\draw [red, thick, dashed] (4.4, 0.1) -- (4.4, -2.8);
			\node [red] at (4.4, 0.4) {$|\psi_2\rangle$};
			
			hspace {8mm} a;
			hspace {8mm} b;
			hspace {8mm} c;
			hspace {8mm} d;
			
			measure a;
			measure b;
			measure d;
			
			box {classical\\optimizer} (a,b,c,d);
		\end{yquant}
		\draw[->] ([xshift=2.38cm, yshift=-0.9cm]current bounding box.center) -- ([xshift=0.6cm, yshift=-0.9cm]current bounding box.center) node[midway, above] {\small update};
	\end{tikzpicture}
	\caption{Structure of the QNN used to solve the DAE, including a classical optimizer that updates \(\boldsymbol{\theta}\).}
	\label{fig:general_arch}
\end{figure*}

\subsection{Framework for Generalized DAEs}
Equations in both the SMIB and the WSCC 3-machine systems serve as specific examples of differential equations. Let us now generalize the systems of DAE. 

We define the set \(S_t\) as follows:
{\footnotesize $$
	S_t = \left\{t, y_1(t), y'_1(t), \ldots, y^{(m_1)}_1(t), \ldots, y_n(t), y'_n(t), \ldots, y^{(m_n)}_n(t)\right\}$$
}
where:
\begin{itemize}
	\item \( t \) represents the independent variable, typically time.
	\item \( y_i(t) \) denotes the \( i \)-th dependent variable as a function of \( t \).
	\item \( y'_i(t), y''_i(t), \ldots, y^{(m_i)}_i(t) \) represent the first, second, ..., up to the \( m_i \)-th order derivatives of \( y_i(t) \), where \( m_i \) is the highest order of derivative for the \( i \)-th variable.
\end{itemize}

Subsequently, the DAE system can be succinctly expressed as:
\begin{equation}
	\left\{
	\begin{aligned}
		F_1(S_t) &= 0, \\
		F_2(S_t) &= 0, \\
		& \vdots \\
		F_{n_e}(S_t) &= 0,
	\end{aligned}
	\right.
	\label{eq:DAE_generalized_compact}
\end{equation}
where:
\begin{itemize}
	\item \( F_j(S_t) \) symbolizes the functions delineating the interrelations within the DAE system. Each \( F_j \) corresponds to the \( j \)-th equation in the system.
	\item \( n_e \) denotes the number of equations in the DAE system, indicating the size of the system.
\end{itemize}

Boundary conditions are essential for constraining the solution space of DAE systems. They are articulated as a series of equations that the solution must satisfy at predetermined points within the domain, not limited to its boundaries. Formally, these conditions are given by:
\begin{equation}
	\left\{
	\begin{aligned}
		B_1(S_{t_*}) &= 0, \\
		B_2(S_{t_*}) &= 0, \\
		&\vdots \\
		B_{n_b}(S_{t_*}) &= 0,
	\end{aligned}
	\right.
	\label{eq:DAE_boundary_conditions_general}
\end{equation}
where:
\begin{itemize}
	\item \( B_k(S_{t_*}) \) denotes the \( k \)-th boundary condition function, which applies to the system state at a critical point \( t_* \) within the domain.
	\item \( n_b \) represents the total number of boundary conditions enforced on the system.
\end{itemize}

\section{A Unified Quantum Neural Network Framework for DAE}
\label{sec:QNN_for_DAE}


In solving DAEs, our approach employs QNNs that aims to approximate the solutions $y_i(t)$. These QNNs encompass several components – Data Encoding, a Parameterized Ansatz, Measurement, and a Classical Optimizer. Each of these components will be elaborated on in later sections of this paper. The methodology primarily involves iteratively updating the circuit parameters. This process is continued until the error in the computed solution, based on the circuit's output, is reduced to a satisfactory level within the context of the DAE equations. An in-depth illustration of the QNN's architecture, particularly its iterative parameter update mechanism, is presented in Fig. \ref{fig:general_arch}. 


From a high-level perspective, this approach comprises two distinct segments: the quantum segment and the classical segment.
\subsection{Quantum Segment}

To initiate the process, we start with a quantum state represented as:
\begin{equation}
    |\psi_0\rangle = |0\rangle^n
\end{equation}
where \(n\) is the number of qubits in the circuit. We then embed our classical input \(x\) into the circuit through a quantum operator \(E(q(x))\) as expressed in equation (\ref{eq:psi1_abstract}).
\begin{equation}
	|\psi_1(x, q)\rangle = E(q(x))|\psi_0\rangle
	\label{eq:psi1_abstract}
\end{equation}

The encoding function \(q(x)\) serves as a preprocessing step for the classical input before it is embedded into the quantum circuit. Next, a parametric multi-layer quantum ansatz operator \( U(\boldsymbol{\theta}) \) is applied to make the circuit trainable.
\begin{equation}
	|\psi_2(x, q, \boldsymbol{\theta})\rangle = U(\boldsymbol{\theta})|\psi_1(x, q)\rangle
\end{equation}
The \( \boldsymbol{\theta} \) is a matrix of parameters and can be written as
\begin{equation}
	\boldsymbol{\theta} = \begin{bmatrix}
		\boldsymbol{\theta}_{00} & \boldsymbol{\theta}_{01} & \hdots & \boldsymbol{\theta}_{0L} \\
		\boldsymbol{\theta}_{10} & \boldsymbol{\theta}_{11} & \hdots & \boldsymbol{\theta}_{1L} \\
		\vdots    & \vdots & \vdots & \vdots \\
		\boldsymbol{\theta}_{n0} & \boldsymbol{\theta}_{n1} & \hdots  & \boldsymbol{\theta}_{nL}
	\end{bmatrix}\text{,}
\end{equation} 
where \( \boldsymbol{\theta}_{ij} \) represents the parameter in the \( i^{th} \) qubit and in the \( j^{th} \) layer. 

\subsection{Classical Segment}
The solution \( f(x, q, \boldsymbol{\theta}) \) is considered a function of the expectation value of an observable \( \hat{O} \) from the final state of the circuit ($|\psi_2\rangle$). Mathematically, this is given by:
\begin{equation}
	M(x, q, \boldsymbol{\theta}) = \langle\psi_2(x, q, \boldsymbol{\theta})|\hat{O}|\psi_2(x, q, \boldsymbol{\theta})\rangle
\end{equation}
\begin{equation}
	f(x, q, \boldsymbol{\theta}) = g(M(x, q, \boldsymbol{\theta}))
	\label{eq:general_f}
\end{equation}
where \( g \) represents the post-measurement function applied to \( M \), approximating the value of the function.
Subsequently, the quantum models are trained by updating and optimizing the parameter $\boldsymbol{\theta}$ with a classical optimizer guided by a loss function. This loss function comprises two parts, providing a measure of the solution's accuracy.

The first part assesses the discrepancy between the estimated and true boundary values. This is quantified using (\ref{eq:sum_err_boundary}), where $n_b$ denotes the total boundary points. 
\begin{equation}
	\label{eq:sum_err_boundary}
	\text{loss}_{b} = \sum_{i=1}^{n_b}(B_i(S_{t_*}) - f^{\text{model}}(S_{t_*}))^2
\end{equation}
Here, $B_i(S_{t_*})$ is derived from the boundary conditions specified by the problem (\ref{eq:DAE_generalized_compact}), while $f^{\text{model}}$ is obtained from the quantum circuit, as described in (\ref{eq:general_f}).

The second part assesses the degree to which the approximated solutions satisfy the system of DAEs (\ref{eq:DAE_generalized_compact}) for $n_p$ training points. It is expressed as:
\begin{equation}
	\label{eq:sum_err_points}
	\text{loss}_{p} = \sum_{i=1}^{n_p}\sum_{j=1}^{n_e}F_j(S_{t_i})
\end{equation}
Here, $n_e$ denotes the number of DAEs in the system.

Finally, the total loss is computed as a weighted sum of the two components, reflecting the relative importance of boundary accuracy and training point adherence. This is expressed as:

\begin{equation}
	\label{eq:total_loss}
	\text{loss}_{total} = \lambda_1\times\text{loss}_{b} + \lambda_2\times\text{loss}_{p}
\end{equation}

In this equation, \(\lambda_1\) and \(\lambda_2\) are tuning coefficients. These coefficients are adjustable parameters that allow for the fine-tuning of the model, enabling prioritization of either boundary accuracy (\(\lambda_1\)) or adherence at the training points (\(\lambda_2\)). The choice of these coefficients depends on the specific requirements and goals of the model, as well as the characteristics of the problem being addressed.


\subsection{Classical Optimizer}
We have explored several classical optimizers. These include Stochastic Gradient Descent (SGD), Adam optimizer, Broyden-Fletcher-Goldfarb-Shanno (BFGS), Limited-Memory BFGS (L-BFGS), and Simultaneous Perturbation Stochastic Approximation (SPSA).

Among these optimizers, BFGS stands out for its fast convergence and robustness in handling non-convex landscapes, making it advantageous in scenarios with saddle points and flat regions \cite{bfgs}. SGD is computationally efficient but may converge slowly \cite{large_stochastic_gradient_descent}. The Adam optimizer offers adaptive learning rates but can exhibit erratic behavior in high-dimensional landscapes \cite{adam}. SPSA, on the other hand, is a gradient-free optimizer suitable for optimizing circuits subject to noise \cite{spsa}.

Given the nature of our problem, we have opted for the BFGS optimizer to enhance the performance of our QNN circuit. Its capability to navigate non-convex landscapes aligns seamlessly with the intricacies and challenges inherent in our problem domain. Nevertheless, it is imperative to underscore that the choice of an optimizer should be contingent upon the specific demands and constraints of the given problem.

For instance, if the hardware introduces noise, the SPSA optimizer might prove more effective. It is crucial to evaluate the unique attributes of the problem and select an optimizer accordingly to ensure optimal performance. In this study, we utilized ideal simulators devoid of noise, and we opted for BFGS based on the elucidated reasons.

\begin{figure*}[!h]
	\centering
	\resizebox{0.8\textwidth}{!}{
		\begin{tikzpicture}[]
			\begin{yquant}[register/separation=3mm]
				qubit {$\ket{0}$} q[2];
				hspace {12.8mm} q[0];
				hspace {12.8mm} q[1];

				
				box {$R_y(x)$} q[1];

				\draw [red, thick, dashed] (0.5, 0.5) -- (0.5, -1.5);
				\node [red] at (0.5, 0.8) {$|\psi_0\rangle$};

				hspace {4.8mm} q[0];
				hspace {11.8mm} q[1];
				
				\draw [red, thick, dashed] (2.99, 0.5) -- (2.99, -1.5);
				\node [red] at (2.99, 0.8) {$|\psi_1\rangle$};
				
				hspace {15.8mm} q[0];
				hspace {4.8mm} q[1];
				
				h q[0];
				hspace {4.8mm} q[0];
				[name=layer1] box {$R_y(\boldsymbol{\theta}_{00})$} q[0];
				box {$R_y(\boldsymbol{\theta}_{10})$} q[1];
				cnot q[1] | q[0];

				hspace {4.8mm} q[0];
				hspace {4.8mm} q[1];
				
				h q[0];
				
				hspace {4.8mm} q[0];
				hspace {4.8mm} q[1];
				
				\draw [red, thick, dashed] (11.3, 0.5) -- (11.3, -1.5);
				\node [red] at (11.3, 0.8) {$|\psi_2\rangle$};

				box {$R_y(\boldsymbol{\theta}_{01})$} q[0];
				box {$R_y(\boldsymbol{\theta}_{11})$} q[1];
				cnot q[1] | q[0];
				h q[0];
				
				hspace {7.8mm} q[0];
				hspace {7.8mm} q[1];
				
				output {$\langle Z \rangle$} q[0];
				
				\draw [dashed,blue,thick] (1.2, 0.4) rectangle (2.7, -1.7);
				\node [blue] at (1.98, -2.0) {Embedding};
				
				\draw [dashed,blue,thick] (3.3, 0.4) rectangle (11.0, -1.7);
				\node[blue] at (7.2, -2.0) {Ansatz $U(\boldsymbol{\theta})$ with 2 layers};
				
			\end{yquant}
			
		\end{tikzpicture}
		
	}
	\caption{SFQ circuit with $R_y$ embedding and two layers}
	\label{fig:qnn_sf_ry}
\end{figure*}
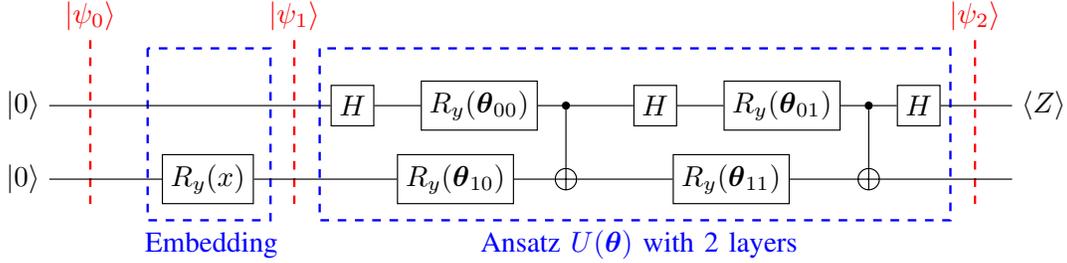

\section{QNN Architectures for Power System Simulation}
\label{sec:QNN_for_simulation}
The implementation of the proposed approach necessitates a pivotal consideration: the selection of a quantum model. This model should not only exhibit expressive capabilities but also closely align with the function it is intended to approximate. In the realm of quantum computing, certain global approximator models, such as Haar Unitary, have been proposed. However, their extensive complexity and the vast parameter space required for training make them less effective. These models often exhibit weak performance, are highly susceptible to encountering saddle point issues, and may become trapped in local minima during the optimization and training process \cite{barren_ansatz,barren_qnn}.

In this paper, we leverage the physics information inherent in power system equations to design quantum models. These models feature generative functions that closely mimic the shapes of power system simulation functions. Consequently, the optimization process becomes significantly more efficient. Power systems involve two distinct types of functions: those characterized by fluctuations and sinusoidal patterns, requiring a model adapted to these features, and those amenable to polynomial fitting, such as power-series and Chebyshev functions.

For sinusoidal-friendly models, we employ our previously proposed Sinusoidal-Friendly QNN (SFQ). Additionally, for polynomial-friendly functions, we introduce a novel and promising quantum model named Polynomial-Friendly QNN (PFQ). PFQ harnesses the power of quantum superposition, offering the potential for quantum advantage.

In the upcoming section, we will delve into the properties of quantum segments for these specific QNNs utilized in our study. 

\subsection{Structure 1 – Sinusoidal Friendly QNN (SFQ)}
\paragraph{Quantum Segment}
The first QNN we utilized is the one we previously explored in \cite{Liao2022} for sinusoidal-friendly functions. We proved that this QNN can efficiently approximate sinusoidal and fluctuating functions. Here, we explain the structure of the circuit we used and leave the details of the expressivity of the model to the paper we mentioned \cite{Liao2022}.

For SFQ, we employed two different kinds of embedding: \( \sin^{-1} \) Embedding (Fig. \ref{fig:qnn_sf_sin^{-1}}), and \( R_y \) Embedding (Fig. \ref{fig:qnn_sf_ry}).

\subsubsection{\( \sin^{-1} \) Embedding} For this type of embedding, we consider two qubits with
\begin{equation}
q(x) = \begin{bmatrix}
    \sin^{-1}(x)  \\
    2\pi x
\end{bmatrix}
\end{equation}
and
\begin{equation}
E(q(x)) = R_y(\sin^{-1}(x)) \otimes R_y(2\pi x)
\end{equation}


\begin{figure*}[!h]
	\centering
	\resizebox{0.8\textwidth}{!}{
		\begin{tikzpicture}[]
			\begin{yquant}[register/separation=3mm]
				qubit {$\ket{0}$} q[2];
				hspace {8.8mm} q[0];
				hspace {8.8mm} q[1];

				[name=embed] box {$R_y(\sin^{-1}(x))$} q[0];
				box {$R_y(2\pi x)$} q[1];

				\draw [red, thick, dashed] (0.5, 0.5) -- (0.5, -1.5);
				\node [red] at (0.5, 0.8) {$|\psi_0\rangle$};

				hspace {4.8mm} q[0];
				hspace {11.8mm} q[1];
				
				\draw [red, thick, dashed] (3.8, 0.5) -- (3.8, -1.5);
				\node [red] at (3.8, 0.8) {$|\psi_1\rangle$};
				
				hspace {4.8mm} q[0];
				hspace {4.8mm} q[1];
				
				h q[0];
				hspace {4.8mm} q[0];
				[name=layer1] box {$R_y(\boldsymbol{\theta}_{00})$} q[0];
				box {$R_y(\boldsymbol{\theta}_{10})$} q[1];
				cnot q[1] | q[0];

				hspace {4.8mm} q[0];
				hspace {4.8mm} q[1];
				
				h q[0];
				
				hspace {4.8mm} q[0];
				hspace {4.8mm} q[1];
				
				\draw [red, thick, dashed] (12.3, 0.5) -- (12.3, -1.5);
				\node [red] at (12.3, 0.8) {$|\psi_2\rangle$};

				box {$R_y(\boldsymbol{\theta}_{01})$} q[0];
				box {$R_y(\boldsymbol{\theta}_{11})$} q[1];
				cnot q[1] | q[0];
				h q[0];
				
				hspace {7.8mm} q[0];
				hspace {7.8mm} q[1];
				
				output {$\langle Z \rangle$} q[0];
				
			   \draw [dashed,blue,thick] (0.8, 0.3) rectangle (3.4, -1.8);
			   \node [blue] at (2.2, -2.1) {Embedding};
			
				\draw [dashed,blue,thick] (11.9, 0.3) rectangle (4.0, -1.8);
				\node[blue] at (8.2, -2.1) {Ansatz $U(\boldsymbol{\theta})$ with 2 layers};
			
			\end{yquant}
			
		\end{tikzpicture}
		
	}
	\caption{SFQ circuit with arcsin embedding and two layers.}
	\label{fig:qnn_sf_sin^{-1}}
\end{figure*}
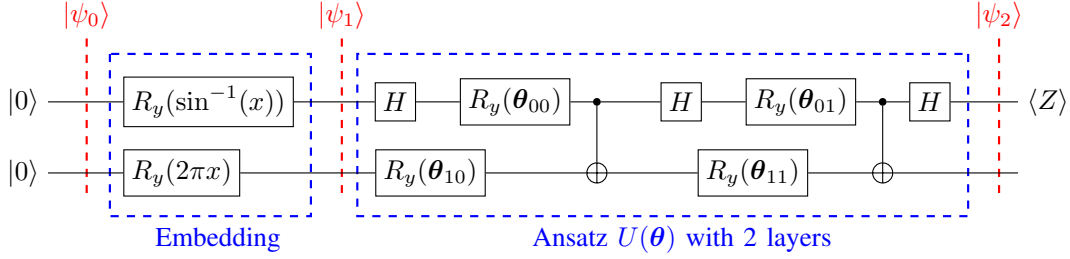

\subsubsection{\( R_y \) Embedding} 
For the \( R_y \) embedding in the SFQ structure, we also consider two qubits but this time
\begin{equation}
q(x) = \begin{bmatrix}
     I \\
    x
\end{bmatrix}
\end{equation}
and, 
\begin{equation}
E(q(x)) = I \otimes R_y(x)
\end{equation}


For implementing the quantum ansatz $U(\boldsymbol{\theta})$, we used a series of layers to train the circuit, fitting it to approximate the solution of our DAE. The ansatz consists of \( L \) layers, and the flexibility and variance of our circuit depend on \( L \) \cite{Liao2022}. The more layers, the more time and data points we need to train it, so the demand on the number of layers depends on the complexity of our function to approximate \cite{Liao2022}. 

The observable \( \hat{O} \) for this structure is \( HZ \otimes I \), where \( H \) is Hadamard gate, \( Z \) is the Pauli-Z matrix, and \( I \) is the identity matrix. The post-processing function \( g \) in this structure is \cite{Liao2022}:
\begin{equation}
	g(u) = \tau_0 + \tau_1 \cdot u + \tau_2 \cdot u^2.
\end{equation}
Assuming that the expectation value of our observable is \( \langle Z_0 \rangle \), then we have
\begin{equation}
	f(x,q,\boldsymbol{\theta}) =  \tau_0 + \tau_1 \cdot \langle Z_0 \rangle + \tau_2 \cdot {\langle Z_0 \rangle}^2\text{.}
\end{equation}

\subsection{Structure 2 – Polynomial Friendly QNN (PFQ)}

We propose an innovative model harnessing quantum principles in QNNs, which exhibit exceptional suitability for approximating functions represented by power series and polynomials. Assuming the ultimate solution of a DAE is expressed as:

\begin{equation}
	h(x) = a_0 + a_1x + a_2x^2 + \ldots + a_nx^m\text{.}
	\label{eq:hx}
\end{equation}

Subsequently, we define $|\psi_0\rangle$ as $$|\psi_0\rangle = |0\rangle^{\otimes{n}}, $$where $n$ is the number of qubits. Afterward, we define the quantum embedding operator $\hat{E}$ as:

\begin{equation}
	|\psi_1(x)\rangle = \hat{E}|0\rangle = \frac{\begin{bmatrix}
			1 & x & \ldots & x^{m}
		\end{bmatrix}^T}{\left\lVert\begin{bmatrix}
			1 & x & \ldots & x^{m}
		\end{bmatrix}^T\right\rVert}\text{,}
	\label{eq:scale}
\end{equation} 
where $m=2^n$.

The construction of the operator $\hat{E}$ is described in \cite{volya2023state} where researchers developed an effective method for quantum state preparation using measurement-induced steering, enabling the initialization of qubits in arbitrary states on quantum computers. Ultimately, a quantum ansatz operator \( U(\boldsymbol{\theta}) \) are utilized to produce a subset of the power series, as expressed in equation (\ref{eq:out0}).
\begin{equation}
	|\psi_2(x,\boldsymbol{\theta})\rangle = U(\boldsymbol{\theta})|\psi_1(x)\rangle
	\label{eq:out0}
\end{equation}

\subsection{Simplified PFQ Form with $m=1$}
To illustrate this process for a single qubit, let's consider an $R_y$ gate as in equation (\ref{eq:out1}).

\begin{equation}
	|\psi_2(x,\theta)\rangle = R_y(\theta)|\psi_1(x)\rangle
	\label{eq:out1}
\end{equation}

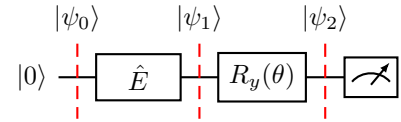
\begin{figure}[H]
	\centering
	\begin{quantikz}
		\lstick{$\ket{0}$}\slice{$|\psi_0\rangle$} & \gate{\hspace{0.8em}\hat{E}\hspace{0.8em}}\slice{$|\psi_1\rangle$} & \gate{R_y(\theta)}\slice{$|\psi_2\rangle$} & \meter  \\ 
	\end{quantikz}
	\caption{PFQ circuit with one rotation gate.}
	\label{fig:pfq1}
\end{figure}

The corresponding quantum circuit with one rotation gate is depicted in Fig.  \ref{fig:pfq1}.

Assuming we set $m$ to 1 in equation (\ref{eq:scale}), we obtain:
\begin{equation}
	\left|\psi_1(x)\right\rangle =  \frac{1}{\sqrt{1+x^2}} \begin{bmatrix}1\\x \end{bmatrix}\text{.}
	\label{eq:d1}
\end{equation}

The $R_y(\theta)$ is defined as:
\begin{equation}
	R_y(\theta) = \begin{bmatrix}\cos{\frac{\theta}{2}}&-\sin{\frac{\theta}{2}}\\\\ \sin{\frac{\theta}{2}}&\cos{\frac{\theta}{2}}\end{bmatrix} \text{.}
	\label{eq:ry}
\end{equation}

By substituting equations (\ref{eq:d1}) and (\ref{eq:ry}) into equation (\ref{eq:out1}), we obtain:

\begin{equation}
	\left|\psi_2(x,\theta)\right\rangle = \frac{1}{\sqrt{1+x^2}} \begin{bmatrix}\cos{\frac{\theta}{2}}-x\sin{\frac{\theta}{2}}\\\\ \sin{\frac{\theta}{2}}+x\cos{\frac{\theta}{2}}\end{bmatrix} \text{.}
	\label{eq:drive}
\end{equation}

Now, let's calculate the probability \(P(|1\rangle)\):
\begin{equation*}
	P(|1\rangle) = \frac{1}{1+x^2}(\sin^2\frac{\theta}{2}+x^2\cos^2{\frac{\theta}{2}+2x\sin{\frac{\theta}{2}}\cos{\frac{\theta}{2}}}) \text{.}
\end{equation*}

After some trigonometric simplifications on $P(|1\rangle)$, we obtain:
\begin{equation}
	P(|1\rangle) = \frac{1}{1+x^2} (\sin^2{\frac{\theta}{2}} + \sin{\theta}\ x + \cos^2{\frac{\theta}{2}}\ x^2) \text{.}
\end{equation}

Multiplying $P(\left|1\right\rangle)$ by $(1+x^2)$ yields:
\begin{equation}
	(1+x^2)P(|1\rangle) = \sin^2{\frac{\theta}{2}} + \sin{\theta}\ x + \cos^2{\frac{\theta}{2}}\ x^2 \text{.}
	\label{eq:key1}
\end{equation}

Equation (\ref{eq:key1}) can be abstracted as:
\begin{equation}
	h_1(x)= b_0 + b_1 x + b_2 x^2
\end{equation}
where $$b_0 = \sin^2{\frac{\theta}{2}},\ b_1 = \sin{\theta},\ b_2 = \cos^2{\frac{\theta}{2}}.$$

This expression can be viewed as a segment of a quadratic power series. However, it encounters some limitations, which we will address subsequently.

\subsection{Enhanced PFQ Version with $m=1$}
Initially, the limitation of simple PFQ concerns the restricted range of \( b_0 \), specifically \( 0 \leq b_0 \leq 1 \). Secondly, the range of \( b_1 \) is also constrained, within \( -1 \leq b_1 \leq 1 \).

To mitigate these constraints, we propose the incorporation of two classical adjustable parameters, enhancing the adaptability of the approximating function. Consequently, we define the output as:

\[ h_2(x) = \tau_1 \cdot h_1(x) + \tau_0 \]

Here, \( \tau_0 \) stands as a classical trainable parameter, whereas \( \tau_1 \) may either be a physics-informed constant, tailored to the problem's boundary conditions, or a trainable parameter. In cases where the boundary conditions are undefined, opting for a trainable \( \tau_1 \) is advisable.

With these adjustments, we obtain:

$$ b_0' = \tau_0 + \tau_1 \cdot b_0, $$ 
and 
$$ b_1' = \tau_1 \cdot b_1. $$

As a result, the range of \( b_0' \) extends to the entire set of real numbers \( \mathbb{R} \), and \( b_1' \) falls within the interval \( -\tau_1 \leq b_1' \leq \tau_1 \).

The third limitation is that the range of $b_2$ is limited, $\left( \text{i.e., } 0 \leq b_2 \leq 1 \right)$.
With the solution of the second limitation, we mitigate the problem of not having negative coefficients for \(x^2\) since we have added \(\tau_1 b_2\), which can create a negative coefficient for \(x^2\). However, we still need to improve it and make it more expressive to have both negative and positive coefficients for \(b_2\). To do this, we can add some rotation gates to add the degree of freedom for the coefficients. As an example, consider we add an \(R_z\) gate after the \(R_y\) gate. Then we have:

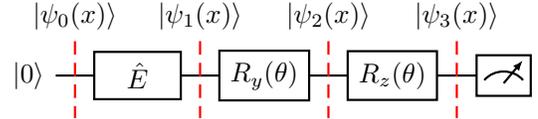
\begin{figure}[!h]
	\centering
	\resizebox{0.40\textwidth}{!}{
		\begin{quantikz}
			\lstick{$\ket{0}$}\slice{$|\psi_0(x)\rangle$} & \gate{\hspace{0.8em}\hat{E}\hspace{0.8em}}\slice{$|\psi_1(x)\rangle$} & \gate{R_y(\theta)}\slice{$|\psi_2(x)\rangle$} & \gate{R_z(\theta)}\slice{$|\psi_3(x)\rangle$} & \meter  \\ 
		\end{quantikz}
	}
	\caption{PFQ circuit with two rotation gates.}
	\label{fig:pfq2}
\end{figure}

\begin{equation}
	\left|\psi_3\left(x,\theta\right)\right\rangle=R_z(\theta)R_y(\theta)\left|\psi_1\left(x\right)\right\rangle=R_z(\theta)\left|\psi_2\left(x,\theta\right)\right\rangle 
	\label{eq:drive2}
\end{equation}

We can write \(R_z\) as:
\begin{equation}
	\small
	R_z(\theta)=\begin{bmatrix}e^{-i\frac{\theta}{2}}&0\\0&e^{i\frac{\theta}{2}}\end{bmatrix}
\end{equation}
or
\begin{equation}
	\small
	R_z(\theta)=\begin{bmatrix}\cos{\frac{\theta}{2}}-i\sin{\frac{\theta}{2}}&0\\\\
		0&\cos{\frac{\theta}{2}}+i\sin{\frac{\theta}{2}}\end{bmatrix}\text{.}
		\label{eq:rz}
\end{equation}

By substituting $|\psi_2(x,\theta)\rangle$ from equation \ref{eq:drive} to \ref{eq:drive2}, we obtain:
\begin{equation}
	\left|\psi_3\left(x,\theta\right)\right\rangle=\frac{1}{\sqrt{1+x^2}}R_z(\theta)  \begin{bmatrix}\cos{\frac{\theta}{2}}-x\sin{\frac{\theta}{2}}\\\\ \sin{\frac{\theta}{2}}+x\cos{\frac{\theta}{2}}\end{bmatrix}
	\label{eq:drive3}
\end{equation}
Consequently, substituting \(R_z(\theta)\) from equation \ref{eq:rz} into \ref{eq:drive3} and multiplying both sides by \(\sqrt{1+x^2}\), yields:

\begin{figure}[H]
	\resizebox{0.8\columnwidth}{!}{
		\begin{minipage}{\columnwidth}
			$$
			\begin{aligned}
				\sqrt{1+x^2} \left|\psi_3(x)\right\rangle &=
				\begin{bmatrix}
					\cos{\frac{\theta}{2}} - i\sin{\frac{\theta}{2}} & 0 \\\\
					0 & \cos{\frac{\theta}{2}} + i\sin{\frac{\theta}{2}}
				\end{bmatrix}
				\begin{bmatrix}
					\cos{\frac{\theta}{2}} - x\sin{\frac{\theta}{2}} \\\\
					\sin{\frac{\theta}{2}} + x\cos{\frac{\theta}{2}}
				\end{bmatrix} \\\\
				&=
				\begin{bmatrix}
					\cos^2{\frac{\theta}{2}} - x\cos{\frac{\theta}{2}}\sin{\frac{\theta}{2}}-i\sin{\frac{\theta}{2}}\cos{\frac{\theta}{2}}+ix\sin^2{\frac{\theta}{2}} \\\\
					\sin{\frac{\theta}{2}}\cos{\frac{\theta}{2}} + x\cos^2{\frac{\theta}{2}} + i\sin^2{\frac{\theta}{2} + ix\sin{\frac{\theta}{2}}\cos{\frac{\theta}{2}}}
				\end{bmatrix} \\\\
				&=
				\begin{bmatrix}
					\cos^2{\frac{\theta}{2}} - \frac{x}{2}\sin{\left(\theta\right) - \frac{i}{2}\sin{\left(\theta\right)} + ix\sin^2{(\frac{\theta}{2})}} \\\\
					\frac{1}{2}\sin(\theta) + x\cos^2{\frac{\theta}{2} + i\sin^2{\frac{\theta}{2} + \frac{1}{2}ix\sin(\theta)}}
				\end{bmatrix}\text{,}
			\end{aligned}
			$$
		\end{minipage}
	}
\end{figure}
which shows that we have \(x \cos^2\left(\frac{\theta}{2}\right)\) and \(ix \sin\left(\theta\right)\) in $\sqrt{1+x^2} \left|\psi_3(x,\theta)\right\rangle$. This implies that when calculating the probability of \(\ket{1}\), terms like \(i \cos^2\left(\frac{\theta}{2}\right) \sin\left(\theta\right) x^2\) can appear, resulting in both negative and positive coefficients of \(x^2\).

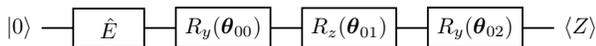
\begin{figure}[!ht]
	\centering
	\resizebox{0.45\textwidth}{!}{
		\begin{quantikz}
			\lstick{$\ket{0}$} & \gate{\hspace{0.8em}\hat{E}\hspace{0.8em}} & \gate{R_y(\boldsymbol{\theta}_{00})} & \gate{R_z(\boldsymbol{\theta}_{01})} & \gate{R_y(\boldsymbol{\theta}_{02})} & \qw \rstick{$\langle Z \rangle$}\\ 
		\end{quantikz}
	}
	\caption{Single-qubit PFQ circuit with three rotation gates.}
	\label{fig:pfq3}
\end{figure}

For simplicity, we assumed the same $\theta$ for both $R_y$ and $R_z$. However, it is also possible to consider different $\theta$ values (e.g., see Fig. \ref{fig:pfq3}) to enhance the model's trainability. Additionally, the introduction of more gates generates additional terms, enabling the model to express a broader range of power series with the same degree.


To evaluate the function to be approximated through this method, we use the same approach as in equation (\ref{eq:general_f}), where $O$ is the expectation value of $Z$ for the first qubit. The expectation value can be calculated as follows:
\begin{equation}
	\langle Z \rangle = P(|0\rangle) - P(|1\rangle)
\end{equation}
Here, we have both $P(|0\rangle)$ and $P(|1\rangle)$, incorporating different coefficients for power series of degree 2.

\subsection{Scalability and Advantage of PFQ}
PFQ's scalability is facilitated by increasing the number of qubits, as outlined in equation (\ref{eq:scale}). Specifically, augmenting the qubit count, $n$, exponentially increases the polynomial's degree, $m$, where $m=2^n$. Furthermore, the introduction of entanglement between qubits, alongside additional rotation gates, enhances the degree of freedom for the coefficients of the generated independent terms in our polynomials. In fact, this approach enables the model to effectively tackle more complex systems by employing extra qubits, thereby extending the degrees of the power series. 

Another critical aspect to note is that the output from our PFQ circuit undergoes post-processing, expressed as
\[
\text{output}(x, \tau_3, \tau_4, \boldsymbol{\theta}) = \tau_3 \times \langle Z \rangle + \tau_4\text{,}
\]
where $\tau_3$ and $\tau_4$ are classical parameters, and $\boldsymbol{\theta}$ represents quantum parameters, with $\langle Z \rangle$ denoting the expectation value of the observable $Z$ for the first qubit. Consequently, the algorithm involves two classical parameters and a flexible number of quantum parameters, enhancing its quantum scalability.


\section{Results}
\label{sec:result}
This section presents the results of using various QNNs to solve DAEs. The QNNs were implemented on Pennylane \cite{bergholm2022pennylane,soltaninia2023comparison} on the NCSA Delta high-performance computer (HPC). The BFGS optimizer was employed to optimize the parameters in the QNNs. Our simulations primarily addressed DAEs in power systems, focusing on 1-machine and 3-machine test systems. Results detail the performance of different QNN structures with different embedding methods (Section \ref{res:QNN_structure}), the setting of time interval and the number of training points (Sections \ref{res:time_interval} and \ref{res:num_training_point}), and the solution of the entire period of the DAE (Section \ref{res:complete_simulation}). 

\subsection{QNN Structure and Embedding}
\label{res:QNN_structure}
\begin{figure}[!h]
	\centering
	\includegraphics[scale=0.45]{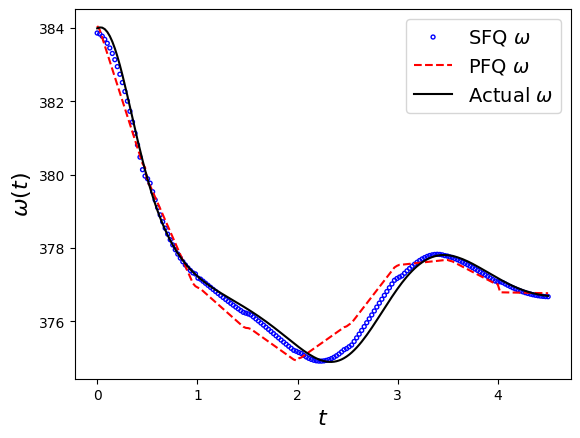}
        \vspace{-8pt}
	\caption{Comparison of actual $\omega$ and the $\omega$ generated by PFQ and SFQ on a 1-machine test system.}
	\label{fig:comparison_structure1}
\end{figure}

To analyze the efficacy of various QNN architectures, we initially focused on a single-machine DE over a 0.5-second interval (the setting of the time interval will be discussed in the next subsection), utilizing 20 training points. The experimental conditions were kept constant, varying only in the QNN structure applied. As depicted in Fig. \ref{fig:comparison_structure1}, the SFQ structure excelled in capturing the sinusoidal behavior inherent in the 1-machine solution, achieving notably lower error compared to the PFQ structure, which exhibited less precision in replicating the solution dynamics.

Transitioning to more complex systems, we investigated the performance of these QNN structures on DAE governing the 3-machine system. This evaluation was conducted over a shorter interval of 0.2 seconds with 10 training points. Our results, as summarized in Table \ref{table:qnn_struct}, demonstrate that all models effectively solve the DAE with minimal error. Note that the $\omega$ and $\delta$ in Tables \ref{table:qnn_struct} and \ref{table:qnn_struct2} represent the rotor speed and rotor angle, respectively.
Notably, the PFQ model, illustrated in Fig. \ref{fig:pfq3}, was particularly adept at managing the complexities of the 3-machine system, displaying significant convergence improvements and an eightfold increase in training speed compared to the SFQ model. 

To further investigate the robustness of these models, we extended the experimental time interval to 2 seconds and increased the number of training points to 50. The expanded results, presented in Table \ref{table:qnn_struct2}, clearly indicate that the PFQ structure outperformed the others. 
It should be noted that due to the restricted domain of the arcsin function, which requires inputs within the range \((-1, 1)\), implementing arcsin embedding within the specified architecture did not successfully solve the equations, resulting in an error, as shown in the 3rd column of Table \ref{table:qnn_struct2}. While rescaling the input through normalization could address this issue, our objective was to maintain the same setup for comparison purposes.

Based on these findings, we suggest selecting the PFQ structure for solving the DE and DAE in power systems and we use it in the rest of the paper, due to its accuracy and shorter training time.


\renewcommand{\arraystretch}{1.5} 
\begin{table}[!ht]
	\centering
\caption{Mean Square Error for the DAE of the 3-Machine WSCC System Between THE Actual Solution and Solutions Obtained by Various QNN Architectures for $0 \leq t \leq 0.2$ (10 Points).}
	\begin{tabular}{lccc}
		\toprule
		& \textbf{SFQ-Ry (Fig. \ref{fig:qnn_sf_ry})} & \textbf{SFQ-Arcsin (Fig. \ref{fig:qnn_sf_sin^{-1}})} & \textbf{PFQ (Fig. \ref{fig:pfq3})} \\
		\midrule
		$\omega_1(t)$ & 1.48e-12  & 1.02e-12 & 9.49e-13  \\
		$\omega_2(t)$ & 1.57e-11  & 5.79e-12 & 9.40e-11 \\
		$\omega_3(t)$ & 9.30e-11  & 1.45e-11 & 5.47e-11  \\
		$\delta_1(t)$ & 1.36e-07 & 1.44e-07 & 1.42e-07 \\
		$\delta_2(t)$ & 4.02e-07  & 3.72e-07 & 3.12e-07 \\
		$\delta_3(t)$ & 9.96e-07  & 1.16e-06 & 1.26e-06 \\
		\midrule
		\textbf{Average} & \(2.56 \times 10^{-7}\) & \(2.79 \times 10^{-7}\) & \(2.86 \times 10^{-7}\) \\
		\bottomrule
	\end{tabular}
	\label{table:qnn_struct}
\end{table}

%
%

\renewcommand{\arraystretch}{1.5} 
\begin{table}[!ht]
	\centering
	\caption{Mean Square Error for the DAE of the 3-Machine WSCC System Between THE Actual Solution and Solutions Obtained by Various QNN Architectures for $0 \leq t \leq 2$ (50 Points).}
	\begin{tabular}{lccc}
		\toprule
		& \textbf{SFQ-Ry (Fig. \ref{fig:qnn_sf_ry})} & \textbf{SFQ-Arcsin (Fig. \ref{fig:qnn_sf_sin^{-1}})} & \textbf{PFQ (Fig. \ref{fig:pfq3})} \\
		\midrule
		$\omega_1(t)$ & 1.40e-09 & failed & 8.62e-10  \\
		$\omega_2(t)$ & 8.15e-09  & failed & 2.11e-09 \\
		$\omega_3(t)$ & 2.43e-07  & failed & 7.31e-09 \\
		$\delta_1(t)$ & 7.76e-06 & failed & 5.20e-07 \\
		$\delta_2(t)$ & 3.92e-04 & failed & 2.37e-06 \\
		$\delta_3(t)$ & 1.27e-02  & failed & 6.67e-05 \\
		\midrule
		\textbf{Average} & \(2.18 \times 10^{-3}\) & - & \(1.16 \times 10^{-5}\) \\
		\bottomrule
	\end{tabular}
	\label{table:qnn_struct2}
\end{table}

%
%

\subsection{Time Interval for DAE Solving}
\label{res:time_interval}
In this paper, the DAE is solved by dividing the time domain into specific intervals, using the final state of one interval as the initial state for the next. This approach simplifies the challenge of function fitting within each interval. 

The selection of an appropriate time interval is critical. A shorter time interval can enhance solution accuracy by limiting the exploration space. However, while a smaller span increases the frequency of DAE resolutions required, it does not necessarily lead to longer overall runtime, as training over shorter spans can be faster, especially for complex models. Determining the best time span involves a trade-off between accuracy and computational efficiency, typically resolved through trial and error.

\begin{figure}[H]
	\centering
    \includegraphics[width=0.33\linewidth]{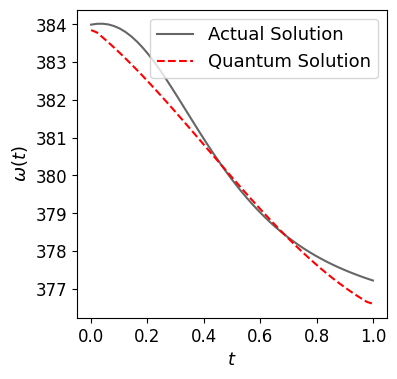}\label{fig:t-1}
    \includegraphics[width=0.31\linewidth]{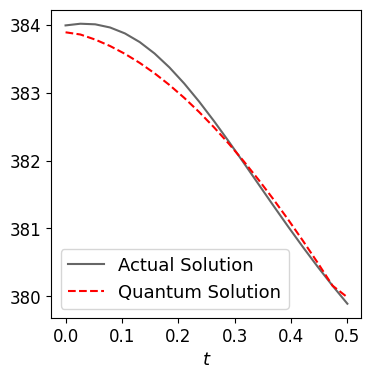}\label{fig:t-0.5}
    \includegraphics[width=0.315\linewidth]{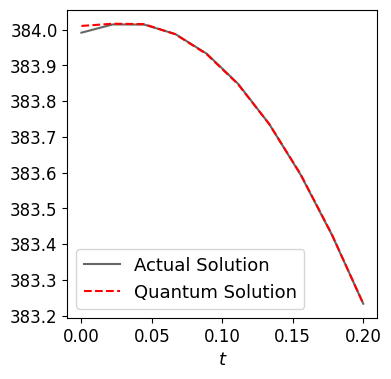}\label{fig:t-0.2}
	\vspace{-8pt}
	\caption{Comparison of quantum and actual solutions for SMIB system using different time spans: left panel $t_s$=1, middle panel $t_s$=0.5, right panel $t_s$=0.2.}
	\label{fig:entire-figure-time}
\end{figure}

Fig. \ref{fig:entire-figure-time} demonstrates the effect of different time spans on the solution accuracy for $\omega(t)$ in the 1-machine system using 20 training points. The results indicate that a time span of $t_s = 0.2$ seconds delivers the most precise outcomes. However, a span of $t_s = 0.5$ seconds also provides satisfactory results, and to minimize the number of computational runs, we selected $t_s = 0.5$ seconds. Following similar experimental guidance, a span of $t_s = 2$ seconds was chosen for the 3-machine DAEs, balancing precision and computational demand.


\subsection{Number of Training Points}
\label{res:num_training_point}
\begin{figure}[H]
	\centering
    \includegraphics[width=0.45\linewidth]{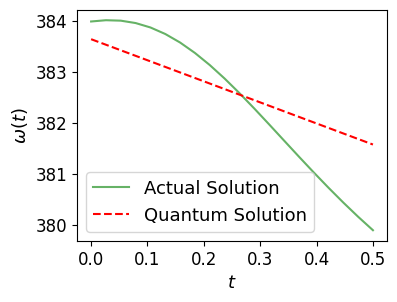}\label{fig:3-pts}
    \includegraphics[width=0.45\linewidth]{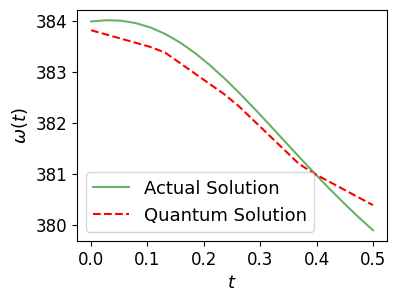}\label{fig:5-pts}
    \includegraphics[width=0.45\linewidth]{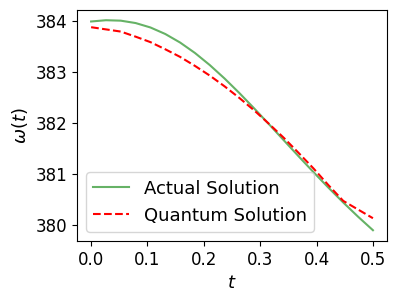}\label{fig:10-pts}
    \includegraphics[width=0.45\linewidth]{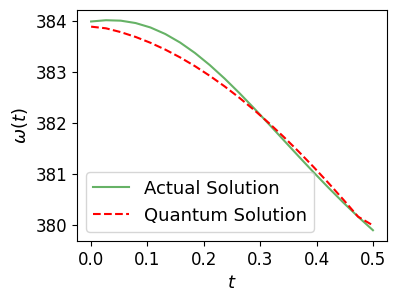}\label{fig:20-pts}
    \vspace{-5pt}	
	\caption{Comparison of quantum and actual solutions for SMIB system using various training points: top left: 3 points, top right: 5 points, bottom left: 10 points, bottom right: 20 points.}
	\label{fig:entire-figure-pts}	
\end{figure}

The number of training points for a quantum circuit is determined by balancing accuracy against computational constraints. Increasing the number of points typically improves precision but also requires more computational time. Hence, optimizing the number of training points involves striking a balance between computational efficiency and desired accuracy. Fig. \ref{fig:entire-figure-pts} illustrates this relationship for the SMIB system, showing how an increased number of training points reduces the error between the quantum approximation and the actual solution for $\omega(t)$ over the interval $t \in [0,0.5]$.



\subsection{Complete Simulation}
\label{res:complete_simulation}
\begin{figure}[H]
	\centering
    \includegraphics[width=0.45\linewidth]{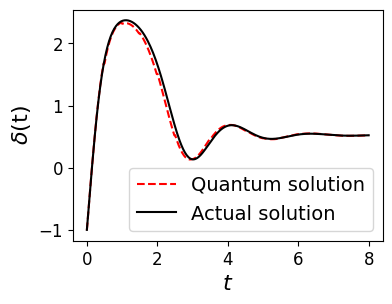}\label{fig:first_case}
    \includegraphics[width=0.46\linewidth]{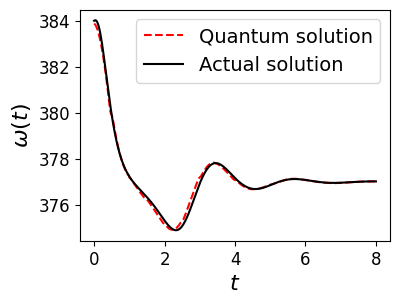}\label{fig:second_case}    
 \vspace{-7pt}
	\caption{Quantum solution vs. actual solution for SMIB system: left panel: rotor angle $\delta(t)$, right panel: rotor speed $\omega(t)$.}
	\label{fig:single-machine}
\end{figure}

By aggregating results over various time intervals, we successfully solved the DAEs for both 1-machine and 3-machine systems. In the 1-machine scenario, we set the time span to 8 seconds with a time interval of \( t_s = 0.5 \) seconds, using 20 training points per interval. We utilized the SFQ structure with \( R_y \) Embedding, and the complete solution is presented in Fig. \ref{fig:single-machine}.

\begin{figure}[H]
	\centering
	\includegraphics[scale=0.38]{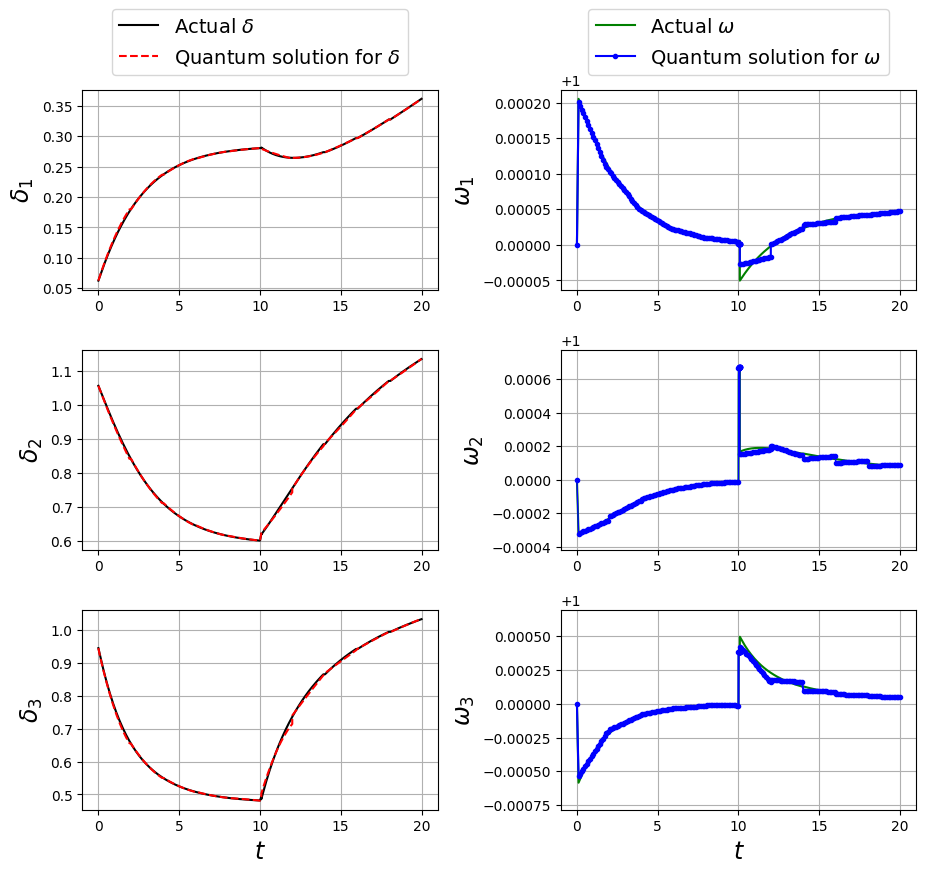}
	\caption{Quantum solution vs. actual solution for WSCC 3-Machine system: the three panels on the left are the rotor angles $\delta$ of the three machines, respectively; the three panels on the right are the rotor speed $\omega$ of the three machines, respectively.}
	\label{fig:3machine_d2}
\end{figure}

For the 3-machine system, we set the time span to 20 seconds, maintaining a time interval of \( t_s = 2 \) seconds, and similarly using 20 training points per interval. The complete solution achieved with the PFQ structure is illustrated in Fig. \ref{fig:3machine_d2}.

The results affirm that the quantum solutions closely align with the actual solutions for both \( \delta \) and \( \omega \) variables, demonstrating the ability of the SFQ and PFQ to accurately simulate the dynamics of power systems in the 1-machine and 3-machine cases, respectively.



\section{Conclusion}
\label{sec:conclusion}
This study demonstrates the potential of Quantum Neural Networks (QNNs) in enhancing the simulation of power system transients by effectively solving differential-algebraic equations (DAEs). The application of QNNs has shown promising results on one-machine and three-machine power systems, capturing the complex dynamics with precision. The advancements in quantum algorithms underscore the potential of quantum computing for practical engineering applications, setting a solid foundation for further research. This work paves the way for broader adoption and scalability of quantum-assisted methodologies in power engineering. Future endeavors will focus on refining these models and expanding their application scope

\section*{Acknowledgments}
This research was supported by the NSF ERI program, under award number 2138702. This work used the Delta system at the National Center for Supercomputing Applications through allocation CIS220136 and CIS240211 from the Advanced Cyberinfrastructure Coordination Ecosystem: Services \& Support (ACCESS) program, which is supported by National Science Foundation grants \#2138259, \#2138286, \#2138307, \#2137603, and \#2138296. We acknowledge the use of IBM Quantum services for this work. The views expressed are those of the authors, and do not reflect the official policy or position of IBM or the IBM Quantum team. We thank I. Tutul for providing the Julia code used to solve the simulation problem of the two systems using classical methods.




%


	\bibliographystyle{IEEEtran}
	
\bibliography{bib/mybibliography,bib/ref_all_Mendeley_2}

\begin{thebibliography}{10}
\providecommand{\url}[1]{#1}
\csname url@samestyle\endcsname
\providecommand{\newblock}{\relax}
\providecommand{\bibinfo}[2]{#2}
\providecommand{\BIBentrySTDinterwordspacing}{\spaceskip=0pt\relax}
\providecommand{\BIBentryALTinterwordstretchfactor}{4}
\providecommand{\BIBentryALTinterwordspacing}{\spaceskip=\fontdimen2\font plus
\BIBentryALTinterwordstretchfactor\fontdimen3\font minus \fontdimen4\font\relax}
\providecommand{\BIBforeignlanguage}[2]{{%
\expandafter\ifx\csname l@#1\endcsname\relax
\typeout{** WARNING: IEEEtran.bst: No hyphenation pattern has been}%
\typeout{** loaded for the language `#1'. Using the pattern for}%
\typeout{** the default language instead.}%
\else
\language=\csname l@#1\endcsname
\fi
#2}}
\providecommand{\BIBdecl}{\relax}
\BIBdecl

\bibitem{Nielsen2011}
M.~A. Nielsen and I.~L. Chuang, \emph{Quantum Computation and Quantum Information: 10th Anniversary Edition}.\hskip 1em plus 0.5em minus 0.4em\relax Cambridge University Press, 2011.

\bibitem{Shor1994}
P.~W. Shor, ``Algorithms for quantum computation: Discrete logarithms and factoring,'' \emph{Proceedings - Annual IEEE Symposium on Foundations of Computer Science, FOCS}, pp. 124--134, 1994.

\bibitem{Grover1996}
L.~K. Grover, ``A fast quantum mechanical algorithm for database search,'' vol. Part F129452, 1996.

\bibitem{Grover1997}
------, ``Quantum mechanics helps in searching for a needle in a haystack,'' \emph{Physical Review Letters}, vol.~79, 1997.

\bibitem{Madsen2022}
\BIBentryALTinterwordspacing
L.~S. Madsen, F.~Laudenbach, M.~F. Askarani, F.~Rortais, T.~Vincent, J.~F. Bulmer, F.~M. Miatto, L.~Neuhaus, L.~G. Helt, M.~J. Collins, A.~E. Lita, T.~Gerrits, S.~W. Nam, V.~D. Vaidya, M.~Menotti, I.~Dhand, Z.~Vernon, N.~Quesada, and J.~Lavoie, ``Quantum computational advantage with a programmable photonic processor,'' \emph{Nature}, vol. 606, pp. 75--81, 6 2022. [Online]. Available: \url{https://www.nature.com/articles/s41586-022-04725-x}
\BIBentrySTDinterwordspacing

\bibitem{Hyyppa2022}
E.~Hyyppä, S.~Kundu, C.~F. Chan, A.~Gunyhó, J.~Hotari, D.~Janzso, K.~Juliusson, O.~Kiuru, J.~Kotilahti, A.~Landra, W.~Liu, F.~Marxer, A.~Mäkinen, J.-L. Orgiazzi, M.~Palma, M.~Savytskyi, F.~Tosto, J.~Tuorila, V.~Vadimov, T.~Li, C.~Ockeloen-Korppi, J.~Heinsoo, K.~Y. Tan, J.~Hassel, M.~Möttönen, Y.~Kuan, J.~Tan, and M.~M. Hassel, ``Unimon qubit,'' \emph{Nature Communications}, vol.~13, pp. 1--14, 11 2022.

\bibitem{Arrazola2021}
J.~M. Arrazola, V.~Bergholm, K.~Brádler, T.~R. Bromley, M.~J. Collins, I.~Dhand, A.~Fumagalli, T.~Gerrits, A.~Goussev, L.~G. Helt, J.~Hundal, T.~Isacsson, R.~B. Israel, J.~Izaac, S.~Jahangiri, R.~Janik, N.~Killoran, S.~P. Kumar, J.~Lavoie, A.~E. Lita, D.~H. Mahler, M.~Menotti, B.~Morrison, S.~W. Nam, L.~Neuhaus, H.~Y. Qi, N.~Quesada, A.~Repingon, K.~K. Sabapathy, M.~Schuld, D.~Su, J.~Swinarton, A.~Száva, K.~Tan, P.~Tan, V.~D. Vaidya, Z.~Vernon, Z.~Zabaneh, and Y.~Zhang, ``Quantum circuits with many photons on a programmable nanophotonic chip,'' \emph{Nature}, vol. 591, 2021.

\bibitem{Lloyd2020NDE}
S.~Lloyd, G.~D. Palma, C.~Gokler, B.~Kiani, Z.-W. Liu, M.~Marvian, F.~Tennie, and T.~Palmer, ``Quantum algorithm for nonlinear differential equations,'' 2020.

\bibitem{JordanZoo}
\BIBentryALTinterwordspacing
S.~Jordan, ``Quantum algorithm zoo.'' [Online]. Available: \url{https://quantumalgorithmzoo.org/}
\BIBentrySTDinterwordspacing

\bibitem{Bharti2021}
K.~Bharti, A.~Cervera-Lierta, T.~H. Kyaw, T.~Haug, S.~Alperin-Lea, A.~Anand, M.~Degroote, H.~Heimonen, J.~S. Kottmann, T.~Menke, W.-K. Mok, S.~Sim, L.-C. Kwek, and A.~Aspuru-Guzik, ``Noisy intermediate-scale quantum (nisq) algorithms,'' 2021.

\bibitem{Zhan2022VQS}
\BIBentryALTinterwordspacing
J.~Zhan, ``Variational quantum search with shallow depth for unstructured database search,'' 12 2022. [Online]. Available: \url{https://arxiv.org/abs/2212.09505v2}
\BIBentrySTDinterwordspacing

\bibitem{Biamonte2017}
J.~Biamonte, P.~Wittek, N.~Pancotti, P.~Rebentrost, N.~Wiebe, and S.~Lloyd, ``Quantum machine learning,'' pp. 195--202, 9 2017.

\bibitem{Liao2022}
\BIBentryALTinterwordspacing
Y.~Liao and J.~Zhan, ``Expressibility-enhancing strategies for quantum neural networks,'' 11 2022. [Online]. Available: \url{https://arxiv.org/abs/2211.12670v1}
\BIBentrySTDinterwordspacing

\bibitem{Skolik2021}
A.~Skolik, J.~R. McClean, M.~Mohseni, P.~van~der Smagt, and M.~Leib, ``Layerwise learning for quantum neural networks,'' \emph{Quantum Machine Intelligence}, vol.~3, 2021.

\bibitem{Feng2021PF}
F.~Feng, Y.~Zhou, and P.~Zhang, ``Quantum power flow,'' \emph{IEEE Transactions on Power Systems}, vol.~36, no.~4, pp. 3810--3812, 2021.

\bibitem{Feng2023UC}
F.~Feng, P.~Zhang, M.~A. Bragin, and Y.~Zhou, ``Novel resolution of unit commitment problems through quantum surrogate lagrangian relaxation,'' \emph{IEEE Transactions on Power Systems}, vol.~38, no.~3, pp. 2460--2471, 2023.

\bibitem{Nikmehr2023Reliability}
N.~Nikmehr and P.~Zhang, ``Quantum-inspired power system reliability assessment,'' \emph{IEEE Transactions on Power Systems}, vol.~38, no.~4, pp. 3476--3490, 2023.

\bibitem{Yifan2023Noise}
Y.~Zhou and P.~Zhang, ``Noise-resilient quantum machine learning for stability assessment of power systems,'' \emph{IEEE Transactions on Power Systems}, vol.~38, no.~1, pp. 475--487, 2023.

\bibitem{Gao2023Review}
\BIBentryALTinterwordspacing
F.~Gao and G.~Wu, ``Application of quantum computing in power systems,'' \emph{Energies}, vol.~16, no.~5, 2023. [Online]. Available: \url{https://www.mdpi.com/1996-1073/16/5/2240}
\BIBentrySTDinterwordspacing

\bibitem{Golestan2023Review}
\BIBentryALTinterwordspacing
S.~Golestan, M.~Habibi, S.~{Mousazadeh Mousavi}, J.~Guerrero, and J.~Vasquez, ``Quantum computation in power systems: An overview of recent advances,'' \emph{Energy Reports}, vol.~9, pp. 584--596, 2023. [Online]. Available: \url{https://www.sciencedirect.com/science/article/pii/S2352484722025720}
\BIBentrySTDinterwordspacing

\bibitem{Zhou2022Review}
Y.~Zhou, Z.~Tang, N.~Nikmehr, P.~Babahajiani, F.~Feng, T.-C. Wei, H.~Zheng, and P.~Zhang, ``Quantum computing in power systems,'' \emph{iEnergy}, vol.~1, no.~2, pp. 170--187, 2022.

\bibitem{Chow2020PowerSystem}
\BIBentryALTinterwordspacing
J.~H. Chow and J.~J. Sanchez-Gasca, \emph{Power System Modeling, Computation, and Control}.\hskip 1em plus 0.5em minus 0.4em\relax John Wiley \& Sons Ltd, 2020. [Online]. Available: \url{https://doi.org/10.1002/9781119546924}
\BIBentrySTDinterwordspacing

\bibitem{Barret2024}
J.-P. Barret, P.~Bornard, and B.~Meyer, \emph{Modelling and Simulation Techniques for Power System Engineering}.\hskip 1em plus 0.5em minus 0.4em\relax Electricite de France, 2024, only 1 left in stock - order soon. FREE delivery January 19 - 24.

\bibitem{Sun2023book}
K.~Sun, \emph{Power System Simulation Using Semi-Analytical Methods}, 2023.

\bibitem{FanMiao2024}
L.~Fan and Z.~Miao, \emph{Modeling and Stability Analysis of Inverter-Based Resources}.\hskip 1em plus 0.5em minus 0.4em\relax CRC Press, 2024.

\bibitem{DE2}
G.~Wanner and E.~Hairer, \emph{Solving ordinary differential equations II}.\hskip 1em plus 0.5em minus 0.4em\relax Springer Berlin Heidelberg New York, 1996, vol. 375.

\bibitem{Liu2020}
Y.~Liu and K.~Sun, ``Solving power system differential algebraic equations using differential transformation,'' \emph{IEEE Transactions on Power Systems}, vol.~35, pp. 2289--2299, 2020.

\bibitem{pinn}
\BIBentryALTinterwordspacing
C.~Moya and G.~Lin, ``Dae-pinn: a physics-informed neural network model for simulating differential algebraic equations with application to power networks,'' \emph{Neural Computing and Applications}, vol.~35, pp. 3789--3804, 2 2023. [Online]. Available: \url{https://arxiv.org/abs/2109.04304v1}
\BIBentrySTDinterwordspacing

\bibitem{Oz2023}
\BIBentryALTinterwordspacing
F.~Oz, O.~San, and K.~Kara, ``An efficient quantum partial differential equation solver with chebyshev points,'' \emph{Scientific Reports}, vol.~13, no.~1, p. 7767, May 2023. [Online]. Available: \url{https://doi.org/10.1038/s41598-023-34966-3}
\BIBentrySTDinterwordspacing

\bibitem{Sauer2017power}
P.~W. Sauer, M.~A. Pai, and J.~H. Chow, \emph{Power System Dynamics and Stability: With Synchrophasor Measurement and Power System Toolbox}, 2nd~ed.\hskip 1em plus 0.5em minus 0.4em\relax Wiley, 2017.

\bibitem{Wang2021DAE}
\BIBentryALTinterwordspacing
B.~Wang, Y.~Liu, and K.~Sun, ``Power system differential-algebraic equations,'' \emph{arXiv:1512.05185 [cs.SY]}, 2021. [Online]. Available: \url{https://doi.org/10.48550/arXiv.1512.05185}
\BIBentrySTDinterwordspacing

\bibitem{bfgs}
\BIBentryALTinterwordspacing
W.~Zhou and D.~Li, ``A globally convergent bfgs method for nonlinear monotone equations without any merit functions,'' \emph{Mathematics of Computation}, vol.~77, 2008. [Online]. Available: \url{https://www.ams.org/mcom/2008-77-264/S0025-5718-08-02121-2/}
\BIBentrySTDinterwordspacing

\bibitem{large_stochastic_gradient_descent}
L.~Bottou, ``Large-scale machine learning with stochastic gradient descent,'' in \emph{Proceedings of COMPSTAT'2010}, Y.~Lechevallier and G.~Saporta, Eds.\hskip 1em plus 0.5em minus 0.4em\relax Berlin, Heidelberg: Physica-Verlag HD, 2010, pp. 177--186.

\bibitem{adam}
S.~J. Reddi, S.~Kale, and S.~Kumar, ``On the convergence of adam and beyond,'' \emph{6th International Conference on Learning Representations, ICLR 2018 - Conference Track Proceedings}, 2018.

\bibitem{spsa}
\BIBentryALTinterwordspacing
M.~Wiedmann, M.~Hölle, M.~Periyasamy, N.~Meyer, C.~Ufrecht, D.~D. Scherer, A.~Plinge, and C.~Mutschler, ``An empirical comparison of optimizers for quantum machine learning with spsa-based gradients,'' 4 2023. [Online]. Available: \url{https://arxiv.org/abs/2305.00224v1}
\BIBentrySTDinterwordspacing

\bibitem{barren_ansatz}
\BIBentryALTinterwordspacing
Z.~Holmes, K.~Sharma, M.~Cerezo, and P.~J. Coles, ``Connecting ansatz expressibility to gradient magnitudes and barren plateaus,'' \emph{PRX Quantum}, vol.~3, no.~1, Jan. 2022. [Online]. Available: \url{http://dx.doi.org/10.1103/PRXQuantum.3.010313}
\BIBentrySTDinterwordspacing

\bibitem{barren_qnn}
\BIBentryALTinterwordspacing
J.~R. McClean, S.~Boixo, V.~N. Smelyanskiy, R.~Babbush, and H.~Neven, ``Barren plateaus in quantum neural network training landscapes,'' \emph{Nature Communications}, vol.~9, no.~1, Nov. 2018. [Online]. Available: \url{http://dx.doi.org/10.1038/s41467-018-07090-4}
\BIBentrySTDinterwordspacing

\bibitem{volya2023state}
D.~Volya and P.~Mishra, ``State preparation on quantum computers via quantum steering,'' 2023.

\bibitem{bergholm2022pennylane}
V.~Bergholm, J.~Izaac, M.~Schuld, C.~Gogolin, S.~Ahmed, V.~Ajith, M.~S. Alam, G.~Alonso-Linaje, B.~AkashNarayanan, A.~Asadi, J.~M. Arrazola, U.~Azad, S.~Banning, C.~Blank, T.~R. Bromley, B.~A. Cordier, J.~Ceroni, A.~Delgado, O.~D. Matteo, A.~Dusko, T.~Garg, D.~Guala, A.~Hayes, R.~Hill, A.~Ijaz, T.~Isacsson, D.~Ittah, S.~Jahangiri, P.~Jain, E.~Jiang, A.~Khandelwal, K.~Kottmann, R.~A. Lang, C.~Lee, T.~Loke, A.~Lowe, K.~McKiernan, J.~J. Meyer, J.~A. Montañez-Barrera, R.~Moyard, Z.~Niu, L.~J. O'Riordan, S.~Oud, A.~Panigrahi, C.-Y. Park, D.~Polatajko, N.~Quesada, C.~Roberts, N.~Sá, I.~Schoch, B.~Shi, S.~Shu, S.~Sim, A.~Singh, I.~Strandberg, J.~Soni, A.~Száva, S.~Thabet, R.~A. Vargas-Hernández, T.~Vincent, N.~Vitucci, M.~Weber, D.~Wierichs, R.~Wiersema, M.~Willmann, V.~Wong, S.~Zhang, and N.~Killoran, ``Pennylane: Automatic differentiation of hybrid quantum-classical computations,'' 2022.

\bibitem{soltaninia2023comparison}
M.~Soltaninia and J.~Zhan, ``Comparison of quantum simulators for variational quantum search: A benchmark study,'' 2023.

\end{thebibliography}

\vfill

\end{document}